\documentclass[aps,prd,twocolumn,superscriptaddress,showpacs]{revtex4}
\usepackage{graphicx}
\usepackage{dcolumn}
\usepackage{bm}
\usepackage{natbib}
\usepackage{multirow}
\usepackage{epsfig}
\usepackage{amsmath}
\newcommand{\Msun}{M_{\odot}}

\newcommand{\beq}{\begin{equation}}
\newcommand{\eeq}{\end{equation}}
\newcommand{\beqa}{\begin{eqnarray}}
\newcommand{\eeqa}{\end{eqnarray}}

\newcommand{\rmd}{{\rm d}}

\begin{document}

\title{
A robust lower limit on the amplitude 
of matter fluctuations in the universe from 
cluster abundance and weak lensing
} 
\author{Rachel Mandelbaum}\email{rmandelb@ias.edu}\thanks{Hubble Fellow} 
\affiliation{Institute for Advanced Study, Einstein Drive, Princeton, NJ 08540, USA}
\author{Uro\v s Seljak}\email{seljak@itp.uzh.ch} 
\affiliation{Institute for Theoretical Physics, University of Zurich, Zurich Switzerland}
\affiliation{Department of Physics, Princeton University, Princeton NJ 08544, U.S.A.}
\affiliation{International Center for Theoretical Physics, Trieste, Italy}

\date{\today}

\begin{abstract}

Cluster abundance measurements are among 
the most sensitive probes of the amplitude of matter fluctuations 
in the universe, which in turn can help constrain other cosmological 
parameters, like the dark energy equation of state or neutrino mass. 
However, difficulties in calibrating the relation between the cluster
observable  
and halo mass, and the lack of completeness information,
make this technique particularly susceptible to systematic errors. 
Here we argue that a cluster abundance analysis using statistical
weak lensing on the stacked clusters leads to a robust 
lower limit on the amplitude of fluctuations. The method compares
the average weak lensing signal measured around the whole cluster sample  
to a theoretical prediction, assuming that the clusters
occupy the centers of all of the most massive halos above some minimum
mass threshold.  If the amplitude of fluctuations is below a certain
limiting value, there are too few massive clusters in this model and
the theoretical prediction falls below the observations.  
Since 
any effects that modify the model assumptions 
can only further decrease the prediction of the model in the context
of this method, 
the limiting amplitude becomes a robust lower limit.
Here, we apply it to a volume limited sample of 16,000 group/cluster
candidates identified from isolated luminous red galaxies (LRGs)
in the Sloan Digital Sky Survey (SDSS). 
We find 
$\sigma_8 (\Omega_m/0.25)^{0.5}>0.62$  
at the 95\% c.l. after taking into account observational 
errors in the lensing analysis. While this is a
relatively weak constraint, both the  
scatter in the LRG luminosity-halo mass relation  
and the lensing errors are large.   
The constraints could improve considerably in the future with 
more sophisticated cluster identification algorithms and smaller 
errors in the lensing analysis. We argue that the existence of a lower limit  
from cluster abundance is rather general, and demonstrate that 
Malmquist bias 
dominates over  Eddington bias in this type of analyses. 

\end{abstract}

\maketitle

\setcounter{footnote}{0}

\section{Introduction}
The abundance of clusters of galaxies has been long recognized as potentially 
one of the most powerful probes of cosmological parameters. 
The main idea is that the cluster abundance can be related to the 
abundance of dark matter halos, which are compared to the theoretical 
halo mass function predictions. These have
an exponential cutoff at the high mass 
end above the so-called nonlinear mass. 
The nonlinear mass is most 
sensitive to the amplitude of 
fluctuations, usually expressed as $\sigma_8$, 
but also depends on other 
cosmological parameters such as the matter density $\Omega_m$.  
By measuring the cluster abundance evolution with redshift, one
can determine the growth rate of structure. 
Combination with other probes such as the cosmic microwave background, 
supernovae, and galaxy clustering
makes cluster abundance measurement a very powerful probe of such fundamental 
parameters as neutrino mass and dark energy density
\cite{2001ApJ...553..545H,2003NewAR..47..775W,2005PhRvL..95a1302W,2006astro.ph..9591A}.  

Existing cluster surveys identify clusters based on an 
observable property like luminosity.  
One typically identifies a flux or surface brightness limited
sample in a given area based on one of the observables, and follow up
observations determine the cluster redshifts via cluster member
spectroscopy. With redshift information, volume-limited samples 
of constant luminosity threshold  can be defined,
and the cluster volume density determined. 
In X-ray surveys, the total cluster luminosity is used as an observable 
\cite{2005astro.ph..7013H}, although other variables may reduce the 
scatter \cite{2006ApJ...650..128K}.  
In optical surveys, either cluster
richness, total luminosity or central galaxy luminosity is used as the
cluster observable \cite{2007ApJ...655..128G,2007astro.ph..1268K,2006MNRAS.372..758M}.
Finally, Sunyaev-Zel'dovich (SZ, \cite{1972CoASP...4..173S})
surveys use either the total SZ luminosity or the central 
decrement. 
To relate the cluster sample to the halo mass
function, one must connect the observable to the halo mass. 
With X-ray data, the 
halo mass can be determined from the measurement of X-ray temperature and 
intensity profile, which allows one to determine the mass profile
through the equation of hydrostatic equilibrium.  Optical surveys may
use velocity dispersion 
measurements, but recently weak lensing calibration has also become
possible  \cite{2006ApJ...653..954D}. 
The SZ signal may already be a reliable tracer of the
dark matter halo mass, since it is only weakly affected by  
astrophysical complications like gas cooling, feedback, cosmic rays, 
etc. \cite{2006ApJ...648..852H,2006astro.ph.11037P}. 
While this technique is still in its infancy, several surveys 
should transform it into reality in the near future. 

In all cosmological 
applications of cluster abundances, 
an accurate calibration between the observable 
and dark matter mass is required. It is important not just to
determine the mean of this relation (e.g. \cite{2005JCAP...12..001F}), 
but also its scatter \cite{2002ApJ...577..569L,2005PhRvD..72d3006L}
and higher moments. 
Small errors in this calibration may significantly 
affect the final cosmological parameters. For example,  
given the steepness of the mass function, a Gaussian
scatter brings many more
low mass halos into the cluster sample than large mass halos out of the 
sample, increasing the abundance for a fixed cutoff in the
observable. On the other hand,  
if a cluster is truly dark, it 
cannot be observed at all, so the cluster sample is incomplete at a 
given mass threshold, and the observed abundance is below the true value. 
In general, given the steepness of the mass function, even a
small deviation from Gaussianity at the tail of the error distribution can
cause a significant effect, which is very difficult to identify 
with current methods using small subsamples of clusters for which the
mass is determined individually. 

The second problem in cluster abundance analyses is that some mass determination methods may be
unreliable because the assumptions on which they rely may be 
violated. For example, mass determination from X-ray measurements of 
gas intensity and temperature using 
hydrostatic equilibrium assumes that the pressure 
support is thermal, while additional sources of pressure (cosmic rays, magnetic fields, 
bulk motions, etc.) are ignored \cite{2006astro.ph.11037P,2006MNRAS.369.2013R}. 
Other X-ray observables (e.g. the product of the X-ray temperature and
the gas mass within $r_{500}$, or $Y_x$, \cite{2006ApJ...650..128K}) may not require
the assumption of hydrostatic equilibrium, but still must be
calibrated against simulations, which may not include all of the
relevant physics.  It is generally accepted that
gravitational lensing is the most  direct tracer of the halo mass, so
any attempts at deriving reliable constraints from cluster abundances
should ideally be based on lensing. 

The third problem is that the calibration between the observable and
the 
halo mass is usually done on a subsample of clusters, which may not  
be representative. An example is 
the relation between mass and X-ray luminosity, which is 
often calibrated on relaxed clusters for which the hydrostatic equlibrium 
assumption may be valid, while the results are then
applied to the whole cluster sample even though non-relaxed clusters
may not obey this relation. 

These arguments suggest that only if clusters are identified by mass
selection, such as when using the dark matter maps derived from gravitational 
lensing, can one avoid these potential systematic errors. 
Unfortunately, mass selection of clusters is still far away from being
practically implemented.  Also, while in principle a
mass-selected sample lacks the difficulties of the current methods, 
the amount of cosmological information is
limited without redshift information.  If the redshift is sought
afterwards, there is again the possibility that the selected cluster candidate 
lacks bright galaxies, bringing in astrophysical complications.
Moreover, there is considerable scatter between the halo mass derived from 
lensing and the more traditional halo mass estimates
\cite{2001ApJ...547..560M}.  For example, in some cases this mass
selection may result in a random superpositions of filaments rather
than true virialized clusters, so one must be able to quantify the
rate of such occurrences.   Alternatively, several promising self-calibration 
methods were proposed for future large data sets
\cite{2003PhRvD..67h1304H,2004ApJ...613...41M}, and  in some cases
were already applied to existing data sets \cite{2007ApJ...655..128G}, but
it remains to be seen whether these methods can reliably solve all of
the aforementioned problems.  

It is worth asking if there is any robust information that can be 
extracted from the cluster abundances without making additional 
assumptions that increase the susceptability to hidden systematic errors. 
We have established above that to derive a robust constraint, we must 
use gravitational lensing to relate the observable to mass; that this
analysis should use the complete sample of clusters; and 
that the effects of scatter and incompleteness are difficult to 
fully account for. Here, we propose a method that uses lensing
analysis on the entire cluster sample. 
Because the issues of completeness and scatter 
cannot be fully solved without additional assumptions, only an
inequality can be established.   

We propose the following analysis. Given a cluster sample with a known
number density at a given redshift, we compute the halo mass function
for a given cosmology (i.e., amplitude 
of fluctuations assuming a value for $\Omega_m$ and other cosmological 
parameters). From this, we can compute the theoretical prediction for the
mean weak lensing signal around the clusters assuming 
that the cluster positions are at the centers of 
all of the most massive halos predicted by the model (with a minimum
halo mass $M_{min}$ determined by the cluster abundance). 
This is the maximally allowed lensing signal for 
that cosmological model, since placing the clusters at any other 
position can only lower the lensing prediction. From the data,
one can compute the mean weak lensing signal around the clusters 
by stacking the individual signals. Stacking reduces the noise, so that 
the resulting $S/N$ can be high even if it is low 
from an individual cluster. Furthermore, it avoids the problems with
individual cluster lensing mass determination mentioned in
\cite{2001ApJ...547..560M}, since we only require that the stacked
signal give information about the cluster mass in the mean.  As in
\cite{2005MNRAS.362.1451M}, comparison with N-body simulations
suggests that this assumption is valid.  

This signal can then be 
compared to the theoretical predictions. These depend mostly on the 
amplitude of fluctuations: lowering it will reduce the 
number of high mass clusters and the lensing signal amplitude. 
At some point the theoretical predictions fall below the observations. 
Such a cosmological model cannot be allowed,  
since any failure of the assumption that the cluster sample 
corresponds to the centers of the most massive 
halos can only reduce the lensing signal, making 
the discrepancy worse. The amplitude of fluctuations for which the 
theoretical prediction still matches the observations within the 
observational errors is thus a robust lower limit. 

This method can be applied to any cluster sample, no matter how incomplete, 
but the derived lower limit improves if the sample is closer to the 
complete sample of most massive halos. 
This may be a small effect if one is on the exponential tail, where 
a small change in amplitude causes a large change in the abundance.
Note that within the context of this method, we cannot establish when the 
inequality becomes equality.  This would require an assessment of
completeness and  scatter in the observable-mass relation,
which as argued above is very difficult and may not even be possible
if some of the clusters are truly dark. 

In this paper, we apply the method to a cluster sample derived 
from isolated luminous red galaxies (LRGs) in Sloan Digital Sky Survey 
(SDSS), for which the high signal to noise of the lensing signal 
has already been established \cite{2006MNRAS.372..758M}. 
The advantages of this sample
are that it covers an unprecedently
large volume and is essentially volume limited. On the other hand, 
the LRG luminosity is unlikely to be a perfect tracer of cluster mass, 
so with a better tracer the constraints may be improved in the future
(for example using the maxBCG cluster sample presented in \cite{2007astro.ph..1265K}). 

Here we note the cosmological model and units used in this paper. 
All data-related computations assume a flat $\Lambda$CDM universe,
though $\Omega_m$ itself is allowed to vary.  Distances
quoted for 
transverse lens-source separation are comoving (rather than physical)
$h^{-1}$kpc, where $H_0=100\,h$ km$\mathrm{s}^{-1}\,\mathrm{Mpc}^{-1}$.
Likewise, $\Delta\Sigma$ is computed using the expression for
$\Sigma_c^{-1}$ in 
comoving coordinates, where $\Sigma_c$ is the critical surface density that 
depends on the distance ratios \cite{2006MNRAS.372..758M}.  For
the typical lens and source redshifts in this study, $\Sigma_c$
depends on cosmology only at the $0.1$\% level even for extreme
changes in cosmology.  In the distance units
used, $H_0$ scales out of everything, so our results are independent of
this quantity.   When masses are quoted in $M_{\odot}$ rather than
$h^{-1}M_{\odot}$, we have used $h=0.7$.  

\section{Data sample and analysis}

As the basis of the cluster sample we use the spectroscopic sample of
LRGs from the latest release of SDSS, data release 5
\cite{2000AJ....120.1579Y,2006ApJS..162...38A}.  
These are good tracers of massive halos, as suggested by their large
bias \citep{2005ApJ...621...22Z}. Previous lensing analyses \citep{2006MNRAS.372..758M}  
have established that the minimum halo mass is $\sim 5\times 10^{13}
\Msun$ and that there is a correlation between LRG luminosity 
and halo mass, $M\propto L^2$, for masses up to $2\times 10^{14} \Msun$. 
The complete, approximately volume-limited sample consists of
$\sim$40,000 galaxies with $0.15<z<0.35$ and a comoving volume of
$0.44 (\mbox{Gpc}/h)^3$.  
We identify a higher-luminosity subsample of host LRGs using spectroscopic galaxy
counts in cylinders of comoving radius 1 
$h^{-1}$Mpc and line-of-sight length $\Delta v = \pm 1200$
km~s$^{-1}$.  Since we only want to eliminate
galaxies for which there is another, brighter spectroscopic LRG
nearby, we simply require that the LRGs in our sample
either (a) be the only one in the cylinder centered on its position,
or (b) be the brightest in 
the cylinder. This cut, which is weaker than the overly-stringent cut
in \cite{2006MNRAS.372..758M} using a 2 $h^{-1}$Mpc cylinder radius,
eliminates about 7.5\% (rather than 13\%) of the complete sample.  

For the LRG luminosity, we use $r$-band model magnitudes
$k+e$-corrected to $z=0$ as in \cite{2006MNRAS.372..758M} to create
luminosity-threshold samples with $M_r < -22.3$, $<-22.6$, and
$<-23$ containing $15~635$, $5~099$, and $902$ galaxies (respectively).    
Because of scatter in $M(L)$, there is no  
guarantee that this will have the desirable effect of selecting the 
most massive among the halos in the sample. 

LRGs targeted for deeper spectroscopy in SDSS have a very specific
color selection \citep{2001AJ....122.2267E}, and it is unlikely that
only the galaxies that satisfy the LRG criteria can be central
galaxies in clusters above our mass threshold.  Indeed, in
\cite{2007astro.ph..1265K}, only 70\% of SDSS
maxBCG clusters in regions with spectroscopy contain LRGs with
spectra. We quantify the effects of color-related incompleteness using a 
comparison against the SDSS Main galaxy sample, which is purely
flux-limited at a brighter limiting flux.  The highest luminosity
sample from our previous Main 
sample lensing analyses \citep{2005PhRvD..71d3511S}, with Petrosian
$r$-band magnitude 
$k$-corrected to $z=0.1$ satisfying $-22 \ge M_r > -23$ (the L6 sample), overlaps
significantly with the spectroscopic LRG sample in redshift, and thus
is an ideal sample for quantifying the effects of the LRG color cut. 
LRGs constitute 65\% of this brightest Main subsample, so one  
possibility is to supplement the LRGs with the galaxies that are
equally as bright  but not as red.  
The Main sample limiting apparent magnitude is 17.77, so these L6
galaxies appear  
significantly brighter than the complete LRG sample ($r<19.1$), partially
because their maximum redshift  
is lower ($\sim 0.25$ rather than $0.35$).  However, their
lensing-weighted mean redshift, $\langle z\rangle=0.2$, is not that  
different from that of the LRG sample, $\langle z\rangle=0.24$, because 
lensing gives the highest signal for the galaxies that are halfway 
between $z=0$ and the sources at redshift 0.3-0.6. 
The L6 Main sample galaxies were split into the 65\% that
pass the LRG selection and the remaining 35\% that fails it. 
\begin{figure}
\includegraphics[width=3.3in,angle=0]{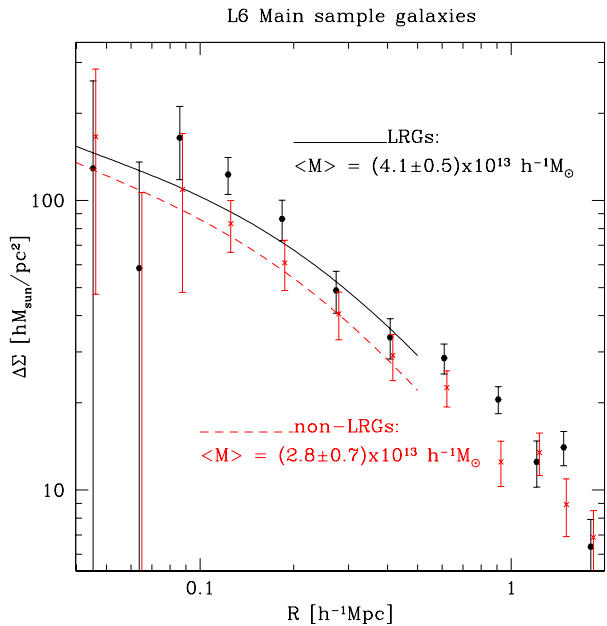}
\caption{\label{F:lrgnotlrg}The lensing signal for the brightest Main sample
  galaxies, divided into those that are spectroscopic LRGs and those
  that are not. We also show the best-fit lensing signal on small
  scales (where the contribution due to host halos for those that are
  satellites is  small),
  giving average estimated virial halo masses using the methodology of
  \protect\cite{2006MNRAS.372..758M}.} 
\end{figure}

As shown in figure~\ref{F:lrgnotlrg}, weak lensing analysis reveals that halos of 
the non-LRG L6 galaxies are $(30 \pm 20)$\% less massive on average.  The
masses were derived from NFW (\citep{1996ApJ...462..563N}) fits to the
inner $0.5h^{-1}$Mpc with fixed concentration $c=6$, where
host group/cluster contributions for satellite lenses are negligible
(since we do not have a reliable environment estimator for L6 that
will allow us to remove satellite galaxies as we do for the LRGs).
The reduced $\chi^2$ values for these fits are $0.53$ (non-LRGs) and $1.0$
(LRGs), which suggests that the NFW model
is an acceptable description of the data (particularly when accounting
for the increase in $\chi^2$ due to noise in the jackknife covariance
matrices, \cite{2004MNRAS.353..529H}).   Unlike in
\cite{2006MNRAS.372..758M}, the virial radius in this case is defined as the radius 
within which the average halo density is $280\overline{\rho}$.  
Allowing the concentration to vary lessens the difference to $\sim
10$\%, as the best-fit concentration for the L6 LRG sample is higher
than that of the L6 non-LRG sample.  
It is important to note that these non-LRG L6 galaxies 
are on average $\sim 0.1$ mag fainter than L6 LRGs.  Given that we have previously
established $M \propto L^2$  at the high-luminosity end
\cite{2006MNRAS.372..758M}, we would indeed expect the L6 non-LRGs to
be 20\% less massive.  Within the errors, these results therefore
justify our assumption that non-LRG galaxies 
are in equally massive halos as LRG galaxies of the same luminosity in
$r$.  Thus, we increase the number density of LRGs by 38\% for $M_r<-22.3$, and
25\% for $M_r<-22.6$ and $M_r<-23$ according to the fractions of L6
galaxies that are classified as LRGs in these bins. We use these increased abundances 
(including the correction for the overly-stringent satellite cut
described previously) in this paper while analyzing the LRG sample 
signal presented in \cite{2006MNRAS.372..758M}.  The abundance corrections due
to the satellite cut vary based on LRG luminosity, and lead to an increase of
7.5\% in the abundance of the complete sample, or 6.0\%, 2.6\%, and 0.7\%
for the three luminosity threshold samples.

The weak lensing analysis is the same as in \cite{2006MNRAS.372..758M}. More than 30 million 
source galaxies are identified, their shape measurements obtained 
using the Reglens pipeline, including PSF
correction done via re-Gaussianization
\citep{2003MNRAS.343..459H} and
with cuts designed to avoid various shear calibration biases.  A full
description of this pipeline can be found in \cite{2005MNRAS.361.1287M}.  
The main difference relative to that paper is that the results of the
Shear TEsting Programme (STEP) 
comparison to simulations are now available and suggest that Reglens
is well calibrated at the 2\% level \cite{2007MNRAS.376...13M}. 
Our redshift distributions are calibrated using the DEEP2 spectroscopic 
survey \cite{2003SPIE.4834..161D,2005MNRAS.361.1287M}. As in
\cite{2005MNRAS.361.1287M}, we assume a total 8\% calibration
uncertainty ($1\sigma$), though this is likely to be a conservative
estimate. 

\begin{figure}
\includegraphics[width=3.3in,angle=0]{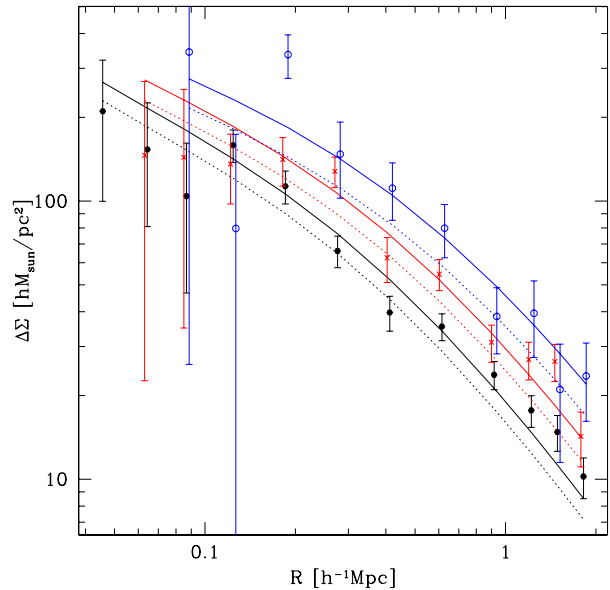}
\caption{\label{F:LRGsignal}The lensing signal for the spectroscopic
  LRGs in luminosity threshold samples, $M_r < -22.3$ (black
  hexagons), $<-22.6$ (red crosses), and $<-23$ (blue circles).
  Bootstrap $1\sigma$ errors are shown, along with the theoretical 
  signal for the best-fit $\sigma_8$ with $\Omega_m=0.25$ (fixed) as a
  solid line, and the $95$\% CL lower limit as a dotted line.}
\end{figure}

Since we know the LRG lens redshifts, we can express the lensing 
signal in terms of the differential surface mass density $\Delta\Sigma$
as a function of transverse separation $R$.  We average the signal over all the clusters, so the result is the 
average weak lensing profile around them. Our lensing signal
estimator, with all associated tests and corrections (for 
imperfect sky subtraction, intrinsic alignments, non-weak shear, and
other effects), is described in
detail in \cite{2006MNRAS.372..758M}.  The results are shown in figure 
\ref{F:LRGsignal} for various luminosity threshold subsamples.   We
see that the signal increases with luminosity, so the luminosity  
thresholds do have the desirable effect of preferentially selecting more 
massive halos. We also see that the signal is inconsistent with zero
at several standard deviations for all luminosity threshold samples
and transverse separations shown in the figure.  The total $S/N$
(over all transverse separations using the full jackknife covariance
matrix) is 19, 17, and 11 for the $M_r<-22.3$, 
$<-22.6$, and $<-23$ threshold samples respectively.  
 
\section{Results}

We want to compare the weak lensing signal to a theoretical prediction 
under the assumption that all of the most massive halos have been 
occupied. We use the Sheth and Tormen halo mass function \cite{1999MNRAS.308..119S}, 
but with parameters as suggested in \cite{2006ApJ...646..881W} and defining the halo 
mass using the radius within which the average density is 
$280\overline{\rho}$.  These two choices represent departures from
the analysis in \cite{2006MNRAS.372..758M}.  
The halo mass function specifies the number density of halos as a 
function of mass, $\rmd n/\rmd M$. For each luminosity threshold we 
select the minimum mass such that the integrated number density equals
the observed one, $\int_{M_{\rm min}}^{\infty}(\rmd n/\rmd M) \rmd M=n_0$, where 
$n_0$ is $5.20$, $1.48$, and $0.26\times 10^{-5} (h/\mbox{Mpc})^3$ for
the three luminosity threshold subsamples (with $\Omega_m=0.25$).  

Once we have the minimum mass, we compute the prediction for the 
weak lensing signal,  
\begin{equation}
\langle \Delta \Sigma \rangle 
=\frac{1}{n_0}\int_{M_{\rm min}}^{\infty}\frac{\rmd n}{\rmd M} \Delta \Sigma (M)
\rmd M,
\end{equation}
where $\Delta \Sigma (M)$ is the differential surface density for 
a cluster of mass $M$, which we model as an NFW profile with 
concentration parameter that depends on the nonlinear mass 
and which must be calculated for each cosmology separately using $c(M)
= 10(M/M_{\rm nl})^{-0.13}$ \cite{2001MNRAS.321..559B}. 
Here $M_{\rm nl}$  is the nonlinear mass, defined 
as the mass in a sphere within which the rms overdensity fluctuation is 1.68. 
The  concentration dependence on the cosmological model mildly  
enhances the scaling with amplitude, because in a model with a lower amplitude 
of fluctuations, not only will $M_{\rm min}$ at a given abundance 
be lower, but so will 
be concentration, which reduces the signal at scales smaller 
than the virial radius.  As in our previous analysis with this sample,
we include a stellar component with mass and radius that is fixed to a
value determined from the optical luminosity. This component is
important for $0.04 < R < 0.08 h^{-1}$Mpc, a small fraction of the
region used for the fits ($0.04<R<2 h^{-1}$Mpc).

\begin{figure}
\includegraphics[width=3.3in,angle=0]{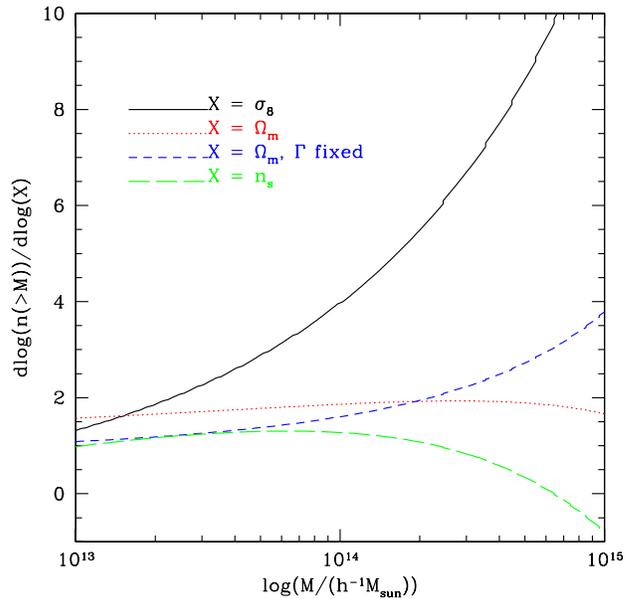}
\caption{\label{F:diffparams}Parameter dependence of the mass function
at the LRG mean redshift for the fiducial cosmology described in the text.}
\end{figure}

This calculation is done separately for each cosmology. 
For the mass function, the most important parameter is the 
nonlinear mass, which depends on the amplitude of fluctuations $\sigma_8$ 
and mass density $\Omega_m$.  For $\sigma_8=0.75$, $\Omega_m=0.25$,
$z=0.24$, $n_s=1$, and $h=0.7$, figure~\ref{F:diffparams} shows the
effect of changing these cosmological parameters on our chosen mass
function and confirms the dominant importance of $\sigma_8$ and $\Omega_m$.  
The shape of the power spectrum also enters the mass 
function calculation \cite{1974ApJ...187..425P}, but the variation in
the mass function 
(figure \ref{F:diffparams})  
is small given the current uncertainties on the shape of the power 
spectrum. For example, the change in 
the mass function 
is three times smaller for $n_s$ than for
$\Omega_m$ when using WMAP3, supernovae, galaxy clustering, and 
Lyman-$\alpha$ forest, \cite{2006JCAP...10..014S}.  Our use of
$n_s=1$ gives conservative bounds on $\sigma_8$, in the sense that
decreasing it by $4$\% to the value from WMAP3 would decrease the
expected abundances by that amount, increasing the best-fit $\sigma_8$
by $\sim 2$\%.  Because it is approximately degenerate with $\Omega_m$
for $M<10^{14}h^{-1}M_{\odot}$, for the fainter threshold samples
where those halo masses dominate, one can estimate the best-fit
$\sigma_8$ for a 4\% lower $n_s$ by simply lowering the fiducial
$\Omega_m$ by that amount instead.

Since we fix $h=0.7$, we
effectively have a different  
power spectrum shape for each value of $\Omega_m$, but as shown in the
figure, the effect of $\Omega_m$ on the nonlinear mass definition is
the main source of our sensitivity to this parameter, so the curves 
of variable $\Omega_m$ with and without fixing the shape parameter 
$\Gamma=\Omega_mh$ are similar.   
Since the matter density affects a number of quantities such as the 
calculation of distances and volumes, the change in 
growth factor from $z=0.24$ to $z=0$, and the definition of virial
radius, we perform the analysis separately for three values, 
$\Omega_m=0.20, 0.25, 0.30$. For each value of $\Omega_m$, we 
vary only $\sigma_8$ while fixing the 
remaining parameters. 

We perform the analysis using $\chi^2$ minimization with respect to 
$\sigma_8$. If $\sigma_8$ is high, then we will have many high 
mass halos, so the predicted lensing signal will exceed the observed
one.  Since deviation from the assumption 
that all of the most massive halos are in our sample can only 
reduce the signal, such a model cannot be excluded. 
Lowering the amplitude of fluctuations reduces the number density 
of massive halos, so 
the lensing signal prediction decreases. As a result, there is a value that fits
the data best. Reducing the amplitude further drops the model
predictions  
below the observations, so such a model cannot be resurrected. 
Figure \ref{F:LRGsignal} shows the results of this 
fitting procedure, showing that we can obtain a good fit for each
luminosity threshold. The best-fit reduced $\chi^2$ values for each luminosity
threshold are $1.4$, $1.2$, and $1.2$ from faintest to
brightest. Accounting for the noise in the covariance matrix, these
$\chi^2$ values give a $p(>\chi^2)$ (probability to exceed the
$\chi^2$ value by chance if the model is a good description of the
data) of $20$\%, $40$\%, and $40$\%.  We also show the prediction of a model which is
excluded  at the 95\% confidence level. 

\begin{figure}
\includegraphics[width=3.3in,angle=0]{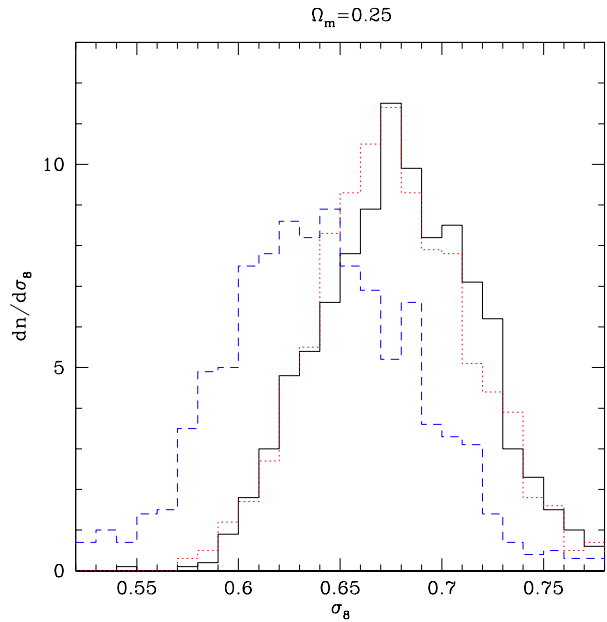}
\caption{\label{F:sigma8dist}The distribution of best-fit $\sigma_8$
  values in the bootstrap subsamples assuming fixed $\Omega_m=0.25$,
  for the three luminosity threshold samples: $M_r < -22.3$ (black,
  solid); $<-22.6$ (red, dotted); $<-23$ (blue, dashed). Each value
should be interpreted as a lower limit due to the failure of modeling
  assumptions. We see that the brightest  
sample gives the weakest limits, suggesting that luminosity fails to trace the 
halo mass with low scatter at the bright end.}
\end{figure}

To determine errors on the lensing signal, we divide the
survey area into 200 bootstrap subregions, and generate 2500
bootstrap-resampled datasets.  
We repeat the analysis for each resampled dataset, introducing the
8\% systematic calibration uncertainty at this stage.  
The final outcome of our analysis is a probability distribution 
for $\sigma_8$ for each luminosity threshold. 
This procedure avoids the problems
\citep{2004MNRAS.353..529H,2007A&A...464..399H} of overly-optimistic 
parameter constraints due to noise in the bootstrap covariance matrix
which tends to increase the fit $\chi^2$ so it deviates from
the usual distribution, and leads to broader parameter distributions
than we would have obtained by naively using $\Delta\chi^2$ values to
find confidence regions.  The distributions are plotted on 
figure \ref{F:sigma8dist} for $\Omega_m=0.25$. We see that the luminosity 
thresholds $M_r < -22.3$ and $<-22.6$ give the strongest constraints on 
$\sigma_8$, while for $M_r<-23$ the constraints weaken.  
If we had a complete sample with no scatter then we should get the 
same constraints for all the samples (assuming the effects from
cosmological parameters besides amplitude are negligible),
so the weaker constraints 
at the bright end suggest that the relation between LRG 
luminosity and halo mass breaks down there. 

We then repeat the analysis for other values of $\Omega_m$, covering the 
range between 0.2 and 0.3. 
We find that we can match the probability distributions by 
introducing a variable $\sigma_8(\Omega_m/0.25)^{0.5}$. 
The final outcome from our analysis is thus the probability 
distribution on this parameter. The constraints can be expressed as
$\sigma_8(\Omega_m/0.25)^{0.5}>0.68$ at 50\% c.l.,
$\sigma_8(\Omega_m/0.25)^{0.5}>0.62$ at 95\% c.l., and
$\sigma_8(\Omega_m/0.25)^{0.5}>0.60$ at 99\% c.l.  
These constraints are for the largest sample, $M_r<-22.3$; we tried
the analysis for the smaller subsamples in case splitting the sample
to reach a higher mass threshold would give superior results,
but unfortunately  this was not the case. 
These constraints can be easily 
implemented in Markov Chain Monte Carlo codes when doing  
global parameter estimates in the form of a soft 
boundary on this parameter combination. 
For the median value of $\sigma_8$ for $\Omega_m=0.25$, the minimum
mass thresholds for the $M_r < -22.3$, $<-22.6$, and $<-23$ samples
are $3.3$, $6.9$, and $14.1\times 10^{13}h^{-1}M_{\odot}$, respectively.

The sources of error for the galaxy-galaxy weak  
lensing analysis have been discussed in the previous section and 
are described in more detail in \cite{2005MNRAS.361.1287M}, and the 
overall calibration uncertainty of 8\% 
has been applied to obtain the final results. 
Similarly, the errors associated with the cluster abundance 
are described in the previous section, and while we apply a correction 
 for the clusters missed by the LRG selection, 
this is correction justified by our lensing analysis of Main sample
luminous galaxies.  
One source of error that we likely do not need to worry about is the 
sampling variance, given that we work with a 
volume of about a cubic Gpc with of order 10,000 clusters. 
For simplicity, we quote the constraint at $z=0$, but the actual 
constraint is at $z=0.24$, and for a given value of $\Omega_m$, 
we have translated between the two 
using a $\Lambda CDM$ cosmological model.
To translate to a different model, for example one with dark energy equation of 
state $w$ different from $-1$, one must 
multiply by the ratio of the growth factors between these two 
redshifts for the two models. Over the range of $w$ allowed by the current 
constraints and the range of redshifts of concern, these effects are
well below the current level of statistical precision.

\section{Discussion}

We have presented a method that can robustly determine a lower limit 
to the amplitude of fluctuations in the universe. 
The method consists of a statistical weak lensing analysis on a
stacked cluster sample compared with theoretical 
predictions, assuming that the sample occupies the 
centers of all of the most massive halos for that model. 
The method does not attempt to establish a mass estimate for each 
cluster, and is thus significantly less observationally demanding than  
the traditional methods of calibration based on a large cluster subsample
for which the mass-observable relation is determined. 
As a result, only an inequality can be established. 
On the other hand, unlike other methods, 
this method uses mass information extracted from 
lensing on the complete cluster sample, removing one of the 
major uncertainties in the traditional approach. 
Furthermore, the main difficulties in the traditional analyses
(knowledge of incompleteness and scatter in the mass-observable
relation) are removed when attempting to place only a lower limit.  
We have
applied the method to a sample of 16,000 cluster candidates from SDSS. 
We find $\sigma_8(\Omega_m/0.25)^{0.5}>0.68$ at 50\% confidence level and 
$\sigma_8(\Omega_m/0.25)^{0.5}>0.62$ at 95\% c.l.,  
where the error is entirely due to statistical and systematic 
errors in the lensing analysis, 
while all modeling 
uncertainties have been absorbed by the inequality. 

At first, it appears surprising that a cluster abundance analysis gives
a robust lower limit. One effect of scatter in the 
mass-observable relation relative to no scatter 
is to increase the number of clusters at a given 
observable threshold, since for a symmetric error distribution around 
the mean, there are always more low mass clusters that scatter 
into the sample than high mass clusters that scatter out of the sample, 
a consequence of the steepness of the halo mass function. 
This increase in abundance is the so-called Eddington bias. 
Ignoring the scatter, one would therefore conclude that 
the derived value of $\sigma_8$ is higher than the true value. 
Consequently, one sometimes finds
statements that, ignoring scatter, the 
cluster abundance gives an upper limit 
on $\sigma_8$ \cite{2006astro.ph..3260P,2006ApJ...653..954D}. 

However, such statements are only valid if the relation between the
mass and observable is extracted from a complete (or at least unbiased) 
sample of halos at a given {\em halo mass} threshold. 
In practice, for a given sample (whether flux- or volume-limited),
this goal is impossible because of scatter: we must 
define the sample used to determine the mass 
observable-relation with a given {\em observable} threshold.
Consequently,  low mass halos for which the observable is above the mean 
will be scattered in, and 
high mass halos for which the observable is below the mean 
will be scattered out.  Since there are more of the former than of the
latter  due to the declining mass
function, the derived mean halo mass at a given observable will 
be underestimated. This is the Malmquist bias effect that acts in 
the opposite direction to the Eddington bias,  
and leads to an underestimate of $\sigma_8$ 
\cite{2006ApJ...648..956S}. 
Note that Malmquist bias is present 
 for both flux-limited and volume-limited samples, 
since the latter are always derived from the former using redshift
information.  
While the two effects oppose each other, our results suggest 
that the Malmquist bias always dominates, so that only a robust lower limit 
can be established. We present an analytical derivation of this result
in Appendix~\ref{A:biases}.  This result should be valid for 
other cluster abundance analyses where the effects of scatter are not 
explicitly taken into account, as long as the halo mass determination 
is reliable (and as long as the sample itself is used to determine the
mass-observable relation, rather than simulations). In practice, mass
determination may not be reliable for those mass 
estimates that are not based on lensing, so one cannot conclude that
all of the derived limits from cluster abundance with no attempt to
correct for scatter should be interpreted as robust lower limits.  
Nonetheless, it is clear that scatter is a serious 
problem, and that it can lead to the opposite effect on $\sigma_8$ 
from what is often assumed. 

Our lower limit on $\sigma_8$ is relatively weak, even in the context of 
the WMAP3 results, which suggest a low amplitude of fluctuations, $\sigma_8 \sim 0.75$ \cite{2006astro.ph..3449S},
 compared to previous analyses giving 0.9 \cite{2005PhRvD..71j3515S}. 
The cosmological constraints we find here 
are likely to be 
improved further with better cluster samples. 
First, the errors in the lensing analysis are significant, and 
the 95\% limit is $0.06$ lower in $\sigma_8$ than the median 50\% 
value, so a larger cluster sample or a deeper survey with more background 
galaxies would reduce the statistical error. 
Second, in our analysis we work with relatively low mass clusters around 
$10^{14}\Msun$ and densities between $10^{-5}$ and $10^{-4}
(h/\mbox{Mpc})^3$. While the mass function is sensitive to the  
amplitude of fluctuations in this range, the sensitivity increases as one moves 
to higher mass clusters with lower number densities, where the mass 
function exponentially decreases, as shown in figure \ref{F:diffparams}. Third, 
using isolated LRG luminosity as a proxy for 
cluster mass is rather unsophisticated.  
While a relation between the halo mass and LRG luminosity has been established
over the range of masses between $5\times 10^{13}\Msun$ 
and $2\times 10^{14}\Msun$, it breaks down above that
\cite{2006MNRAS.372..758M}, so 
one cannot use this method to identify a sample of very massive clusters 
in SDSS. While
the scatter between LRG luminosity and halo mass has not been established, 
it is likely to be significant.  Consequently, 
it is unlikely that the value of the amplitude of fluctuations 
derived in this paper is very close to the true value. 
In Appendix~\ref{A:biases} we derive the amplitude of the effect 
assuming a simple lognormal scatter. The bottom panel of figure~\ref{F:alpha}  
shows that for a 50\% scatter in the mass-observable relation, the 
underestimate of $\sigma_8$ is 6-7\%, relatively independent of the mass
above $10^{14}h^{-1}M_{\odot}$. However, the scatter can be even larger 
and may not follow the simple lognormal model. 
For example, if
we apply corrections to the observed lensing signal consistent with
the results for the brightest luminosity sample in
\cite{2005MNRAS.362.1451M} to account for the effects of scatter in $M(L)$, the
best-fit values of $\sigma_8$ changes by 15\%.  We
do not suggest that the model for $M(L)$ scatter used in that work is
a close enough approximation to the real one that this correction
should be applied; rather, we include this information to make it very
clear that scatter is an important issue for the SDSS LRG sample
that can seriously affect measured cosmological parameters.

In the future, it would be worth repeating this analysis on samples of clusters 
with an even lower number density, for which
the abundance is more sensitive
to the amplitude of fluctuations.  
While the number of clusters will be lower, their lensing signal 
will be higher, so as long as the signal to noise remains high, 
such analysis will yield useful constraints that 
may improve upon the ones established in this paper.  
Many sophisticated algorithms have been developed to 
select clusters and determine their mass based on their richness of 
red galaxies with photometric redshifts \cite{2000AJ....120.2148G}. A particularly promising 
method in the context of the SDSS is MaxBCG \cite{2007astro.ph..1268K}, which identifies clusters
by their concentration of galaxies along the red galaxy color-redshift
relation and 
establishes the mass-richness relation across a broader range of 
halo mass. We expect that applying our method to this and other samples
of clusters 
will yield lower limits on the amplitude that may be a useful complement to the 
traditional analyses of cluster abundances which yield very small formal 
errors, but may contain hidden systematic errors. 

Another possibility 
is to apply this method to the sample of regular 
$L_*$ galaxies, tracing halo masses around $10^{12}\Msun$. 
These galaxies are well below the nonlinear mass, in the regime where the halo 
mass function is sensitive mostly to $\Omega_m(n_{\rm eff}+3)$, where 
$n_{\rm eff}$ is the effective slope of the power spectrum at megaparsec
scales \cite{2002MNRAS.337..774S}. 
Here again only a lower limit to this combination can be established.
However, the errors on these parameters from other
probes are already quite small, and it is not clear
that such analysis would be competitive given the current errors in
the lensing analysis. 

Is it possible that ultimately the lower limit we established here
is all that will 
be  achievable with cluster abundance studies, even when a 
reliable mass tracer such as weak lensing is used? This would be a 
major retreat compared to the expectations 
for future surveys \cite{2001ApJ...553..545H,2005PhRvL..95a1302W,2003NewAR..47..775W}. 
In principle this option cannot be discarded, since there is always 
a possibility that a subset of dark matter halos is not detected 
with X-ray, optical or SZ surveys. More generally, the conditional 
probability distribution of halo mass at a given luminosity may be 
complicated and difficult to establish using subsamples of data. 
While it is likely that further studies, both theoretical and 
observational, will improve our knowledge, the ultimate limitations
of the method
are difficult to establish. It is therefore important to know  
that at least some cosmological information extracted from cluster abundance 
is robust, even if it is only in a form of inequality. 

We thank Chris Hirata for numerous discussions. 
R.M. is supported by NASA
through Hubble Fellowship grant \#HST-HF-01199.02-A awarded by the
Space Telescope Science Institute, which is operated by the
Association of Universities for Research in Astronomy, Inc., for NASA, 
under contract NAS 5-26555.  U.S. is supported by the
Packard Foundation 
and NSF CAREER-0132953.  

\bibliography{cosmo,cosmo_preprints}

\appendix

\section{Analytic comparison of Eddington versus Malmquist
  bias}\label{A:biases} 

Here, we analytically compare the magnitudes of Eddington and
Malmquist biases with minimal assumptions for an analysis that 
proceeds as follows: determining the mass-observable
relation from the lensing signal for the complete sample, and
using the observed abundance with that mass normalization to constrain
$\sigma_8$.  Due to our strict cut
in the observable, the steep decline of the mass
function is going to 
(a) bias the average mass low (Malmquist bias) and (b) bias the
abundance high (Eddington bias), whether
the sample is flux- or volume-limited.  We
will now analytically prove our claim that the former effect dominates
over the latter, leading to an underestimate of $\sigma_8$. 
We demonstrate this effect assuming a lognormal scatter in the
mass-observable relation without placing a requirement on the size of
the scatter.

The parameters of our analysis are the true cluster mass $M$ and the
estimate of the mass $\tilde{M}$ derived from a
cluster observable such as richness, X-ray luminosity, etc.
Due to our use of a lognormal error
distribution (i.e., constant relative rather than absolute scatter in
the observable  as a 
function of mass), our calculation uses the variables $x=\ln{(M/M_0)}$ and
$\tilde{x}=\ln{(\tilde{M}/M_0)}$ for some arbitrary $M_0$.  The relation between $\tilde{x}$
and $x$ is then
\beq
p(\tilde{x}|x) = \frac{1}{\sqrt{2\pi\sigma^2}} \exp{[-(\tilde{x}-x)^2/2\sigma^2]}.
\eeq

We assume the halo mass function can be expressed as a summation over
power laws, so we perform the calculations for a single power law and
determine over what ranges of power law slope the calculation is valid.
When using 
\beq
\frac{dn}{dM} = \frac{n_0}{M_0} \left(\frac{M}{M_0}\right)^\alpha
\eeq
we then find $dn/dx = n_0 \exp{[(\alpha+1)x]}$.

\subsection{Observed abundance calculation}

First, we assume that the mass threshold is placed using the observable by
requiring $\tilde{x} > x_c$ for some $x_c$.  The observed number density is then
\beq\label{E:obsabundance}
n(\tilde{x}>x_c) = \int_{x_c}^{\infty} \rmd\tilde{x} \int_{-\infty}^{\infty}
\rmd x\frac{\rmd n}{\rmd x} p(\tilde{x}|x) = \int_{x_c}^{\infty} \rmd\tilde{x} \frac{\rmd\tilde{n}}{\rmd\tilde{x}} 
\eeq
where $\rmd\tilde{n}/\rmd\tilde{x}$ is the observable mass function 
including effects of scatter.  The integral over $x$ can be performed
by completing the square to obtain 
\beqa
\frac{\rmd\tilde{n}}{\rmd\tilde{x}} &=& n_0 \exp{[(\alpha+1)\tilde{x}]}
\exp{[\sigma^2(\alpha+1)^2/2]} \notag \\
 &=& \frac{\rmd n(\tilde{x})}{\rmd x} \exp{[\sigma^2(\alpha+1)^2/2]}
\eeqa
The latter term represents an increase in abundance due to scatter relative to the
true halo mass function.  Performing the integral over $\tilde{x}$ in
equation~\ref{E:obsabundance} for $\alpha<-1$ (true for all mass
ranges of interest in this paper) yields
\beqa
n(\tilde{x}>x_c) &=&
\frac{n_0}{-(\alpha+1)}\exp{[(\alpha+1)x_c]}\exp{[\sigma^2(\alpha+1)^2/2]}
\notag \\
 &=& n(x>x_c)\exp{[\sigma^2(\alpha+1)^2/2]}\label{E:obsabundance2}
\eeqa
The last factor in this equation represents the Eddington bias (which is 
always greater than unity) and shows 
that the observed abundance is higher than it would have been 
at the same threshold in the absence of scatter. 
This effect 
is stronger for a steeper mass function because, for a fixed scatter,
there are even more 
lower mass halos to scatter into the sample when the mass function is steeper.
If the scatter is ignored then this effect causes one to 
overestimate $\sigma_8$ from the observed abundance. 

\subsection{True mean mass}

We now compute the true mean mass of the sample, which must also be
affected by the lower mass halos scattering into the sample.  This is
the mass that will be measured from the sample 
assuming the mass estimator is correct.  
Since we have assumed that the scatter in mass is lognormal, 
i.e. gaussian in $x$, 
we compute the average of $x$ rather than of the mass itself, 
\beq
\langle x\rangle = \frac{1}{n(\tilde{x}>x_c)}\int_{x_c}^{\infty}
\rmd\tilde{x} \int_{-\infty}^{+\infty} x \frac{\rmd n}{\rmd
  x}p(\tilde{x}|x) \rmd x.
\eeq
The integral over $x$ can again be performed by completing the
square, and after carrying out the integral over $\tilde{x}$ we obtain
\beq\label{E:meanmass}
\langle x \rangle = x_c - \frac{1}{\alpha+1} + (\alpha+1)\sigma^2=\langle \tilde{x} \rangle+ (\alpha+1)\sigma^2.
\eeq
The first term $\langle \tilde{x} \rangle$ is the estimated mean mass 
assuming our mass estimator. It
equals the mean true mass in the absence of
scatter.
The final term represents the Malmquist bias; 
because $\alpha+1<0$, scatter causes the estimated mean mass to be biased
high relative to the true mean mass. Thus, if one neglected scatter, 
one would conclude from this sample that the mass estimator is biased 
and would apply a correction from equation \ref{E:meanmass} to rectify it, 
even though we have modeled the scatter between the estimated and 
true value of $x$ as a gaussian with zero mean. 
In the absence of an additional external calibration this is the only 
way to calibrate the estimator. 
While we have assumed that the mass estimator is given for every cluster 
in the sample, a random subsample of the complete sample would lead to the same 
result. 

\subsubsection{Effects of ignoring scatter}

We now consider the competition between the computed Malmquist and
Eddington biases in an analysis that ignores
the effects of scatter.  From lensing, we see a mean mass given by
equation~\ref{E:meanmass} and determine an underestimated lower mass
cutoff for our sample,
\beq
\hat{x}_c = x_c + (\alpha+1)\sigma^2 < x_c.
\eeq
Next, we compute the expected abundance for that lower mass cutoff
assuming no scatter in the 
mass-observable relation, i.e.
\beqa
n(x>\hat{x}_c) &=& \int_{\hat{x}_c}^{\infty} \rmd x \frac{\rmd n}{\rmd
  x} \notag \\
 &=& \frac{n_0}{-(\alpha+1)} \exp{[(\alpha+1)\hat{x}_c]} \notag \\
 &=& n(x>x_c)\exp{[(\alpha+1)^2\sigma^2]} \label{E:expectedn}
\eeqa

Finally, we compare the true observed abundance including 
scatter, equation~\ref{E:obsabundance2}, against the expected
abundance in our analysis without scatter, 
equation~\ref{E:expectedn}.  We find
\beqa
\frac{n(\tilde{x}>x_c)}{n(x>\hat{x}_c)} &=& \frac{\exp{[\sigma^2(\alpha+1)^2/2]}}{\exp{[\sigma^2(\alpha+1)^2]}} \notag \\
 &=& \exp[-\sigma^2(\alpha+1)^2/2] < 1.
\eeqa

\begin{figure}
\includegraphics[width=3.3in,angle=0]{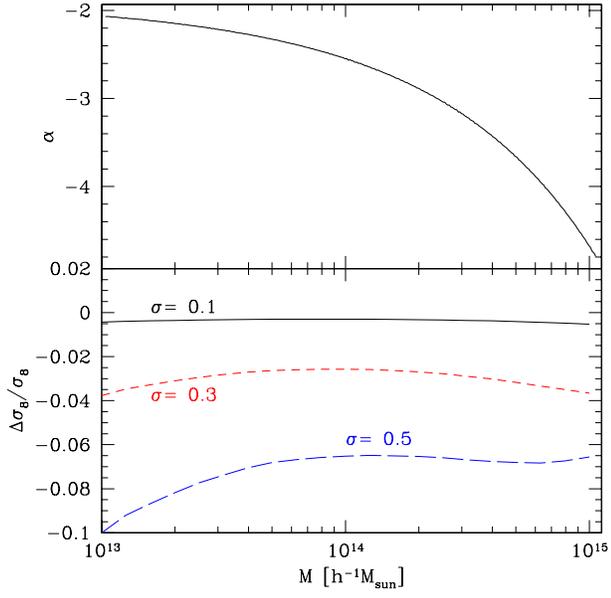}
\caption{\label{F:alpha}Top panel: power law index of $\rmd n/\rmd M$
  for the mass range of interest for this work.  Bottom panel:
  resulting bias in $\sigma_8$ when ignoring scatter in the
  mass-observable relation at the 10\%, 30\%, and 50\% level.}
\end{figure}

Thus, for an arbitrary size lognormal scatter in the
mass-observable relation, the Malmquist bias dominates over
the Eddington bias, leading us to underestimate $\sigma_8$.  
This conclusion is valid when the mass function can be expressed
locally (near $x_c$) as a
power law in mass with slope steeper than $-1$, which is true for all
masses of interest in this work and, indeed, in most typical cluster
abundance analysis.  As a demonstration, figure~\ref{F:alpha} shows
$\alpha$ (the power-law index of $\rmd n/\rmd M$) for $\Omega_m=0.25$,
$\sigma_8=0.75$, $z=0.25$, and $n_s=1.0$ 
for the mass ranges of interest, for which $\alpha<-2$.  The bottom
panel also shows the expected bias in $\sigma_8$ as a function of
lower mass cutoff for fixed
$\sigma=0.1$, $0.3$, and $0.5$ (no variation with mass) given the
$\alpha$ from the top panel and the change in predicted abundance with
$\sigma_8$ from figure~\ref{F:diffparams}.  This bias arises due to
the dominance of Malmquist bias leading to an apparent underestimate
of number density, which is interpreted in an analysis without scatter
as a suppression of $\sigma_8$.

\end{document}